\documentclass[%
 aps,
 amsmath,amssymb,
 reprint,%
]{revtex4-1}
\usepackage{latexsym}
\usepackage{hyperref}
\usepackage{graphicx}%
\usepackage{xcolor}
\usepackage{subfigure} 
\usepackage{dcolumn}%
\usepackage{bm}%
\usepackage{upgreek}
\draft 
\usepackage[utf8]{inputenc}
\usepackage[T1]{fontenc}
\usepackage{mathptmx}
\usepackage{float}
\begin{document}


\title{A unified ballistic transport relation for anisotropic dispersions and generalized dimensions}

\author{Jashan Singhal}
\email{js3452@cornell.edu}
\affiliation{School of Electrical and Computer Engineering, Cornell University, Ithaca, New York 14853, USA,}
\author{Debdeep Jena}%
\affiliation{School of Electrical and Computer Engineering, Cornell University, Ithaca, New York 14853, USA,}
\affiliation{Department of Materials Science and Engineering, Cornell University, Ithaca, New York 14853, USA}
\affiliation{Kavli Institute at Cornell for Nanoscale Science, Cornell University, Ithaca, New York 14853, USA}





\date{\today}

\begin{abstract}
An analytical formula is derived for particle and energy densities of fermions and bosons, and their ballistic momentum and energy currents for anisotropic energy dispersions in generalized dimensions. The formulation considerably simplifies the comparison of the statistical properties and ballistic particle and energy transport currents of electrons, acoustic phonons, and photons in various dimensions in a unified manner. Assorted examples of its utility are discussed, ranging from blackbody radiation to Schottky diodes and ballistic transistors, quantized electrical and thermal conductance, generalized ballistic Seebeck and Peltier coefficients, their Onsager relations, the generalized Wiedemann-Franz law and the robustness of the Lorenz number, and ballistic thermoelectric power factors, all of which are obtained from the single formula. The new formulation predicts a thermoelectric power factor behaviour of 3D Dirac bands which has not been observed yet. 
\end{abstract}

\maketitle 

{\bf Introduction:} The need for analytical expressions for particle, energy, and current densities arises frequently in various branches of science and engineering.  They are typically handled separately for each case of interest. This is because the densities depend on the quantum statistics of the type of particle or field of interest (i.e., whether they are fermions or bosons), on their specific energy dispersions (e.g. $E = \hbar^2 \lvert\mathbf{k}\rvert^2/2m$ or $E=\hbar v_F | \mathbf{k} | $), or the specific dimensionality under consideration (e.g. $d=1,2,3$). A single unified analytical expression is found in this work for all the above densities and their ballistic momentum and energy currents for anisotropic dispersions in the  non-interacting ballistic transport regime. This enables particle and energy densities, and ballistic particle and energy transport currents of electrons, phonons and photons to be treated in a unified manner amplifying their similarities and differences, the need for which has been advocated \cite{chen2005nanoscale}. Though the discussion in this work is limited to electrons, phonons and photons, the results apply to  ballistic transport in general, such as that of ultra-cold atoms and molecular gases (e.g. \cite{keerthi2018ballistic}).

{\bf Setup:} For particles in a box of dimension $d = 1, 2, 3$ and volume $L^d$, wave-particle duality allows discrete wavevectors $k_i = p_i (2\pi/L)$ where $p_i=0, \pm1, ...$ are integers. The resulting energy dispersion is written as $E = [ \sum_{i=1}^{d} (\alpha_{i}  k_i)^2 ]^{\frac{t}{2}}$. Here the {\em type} $t=1$ represents linear (or conical) dispersion with $\alpha_i = \hbar v_i$ and $t=2$ represents parabolic dispersion with $\alpha_i = \hbar/\sqrt{2m_i}$, where $\hbar=h/2\pi$ is the reduced Planck's constant.  Table \ref{tab:dispersions} shows that this formulation captures anisotropic dispersions via direction-dependent wave velocities $v_i$ (e.g. anisotropic, non-dispersive and transparent optical or acoustic media) or effective masses $m_i$ (e.g. the electron energy bandstructure of the semiconductor Silicon). Though the table and the following discussion is restricted to massless Dirac-like and massive parabolic dispersions, the formulation holds for other $t$. Extensions to other dispersions ought to be feasible along similar lines. 

\begin{table*}
\caption{\label{tab:dispersions}Generalized energy dispersion in $d-$dimensions}
\begin{ruledtabular}
\begin{tabular}{cccccc}
\multicolumn{6}{c}{$E = [ \sum_{i=1}^{d} ( \alpha_{i} k_i)^2 ]^{\frac{t}{2}}$ }\\ \hline 
& $t$ & $\alpha_i$ & $d=1$&$d=2$&$d=3$\\ \hline
 &   &  &  &  &  \\
Conical & $1$  & $ \hbar v_{i}$ & $\hbar v_F k_1$ & $\hbar \sqrt{ (v_1 k_1)^2 + (v_2 k_2)^2 }$ & $\hbar \sqrt{ (v_1 k_1)^2 + (v_2 k_2)^2 + (v_3 k_3)^2 }$ \\
 &   &  &  &  &  \\
Parabolic & $2$  & $\frac{\hbar}{\sqrt{2 m_i}}$ & $\frac{\hbar^2 k_1^2}{2 m_1}$ & $\frac{\hbar^2 k_1^2}{2 m_1} + \frac{\hbar^2 k_2^2}{2 m_2}$ & $\frac{\hbar^2 k_1^2}{2 m_1} + \frac{\hbar^2 k_2^2}{2 m_2} + \frac{\hbar^2 k_3^2}{2 m_3}$ \\
\end{tabular}
\end{ruledtabular}
\end{table*}

\begin{figure}[!htbp]
\includegraphics[width = 0.5\textwidth]{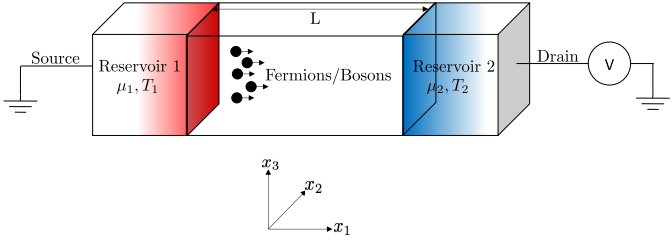}
\caption{\label{3dmodel} Fermionic or Bosonic systems whose ballistic transport is explored in this work. The `particles' may be electrons, photons, phonons, or atoms or molecules, in a potential that produces either a parabolic or conical energy eigenvalue dispersion with momentum.}
\end{figure}

`Source' (1) and `drain' (2) reservoirs, characterized by dimensionless parameters $\eta_1 = \beta_1 \mu_1$ and $\eta_2 = \beta_2 \mu_2$ are connected to the box of particles of dimension $d$ on opposite faces of dimension $d-1$ as shown in Fig.~\ref{3dmodel}.  Here $\beta=(k_b T)^{-1}$ where $k_b$ is the Boltzmann constant. The chemical potentials $\mu_1$ and $\mu_2$ and temperatures $T_1$ and $T_2$ of the source and drain may in general be different.  The particles in the source and drain reservoirs follow the equilibrium distribution functions $f_{\pm}(E) = 1/(\exp[\beta(E-\mu)] \pm 1)$ with + for fermions and - for bosons with the corresponding chemical potential and temperature.  

The particles in the box are in quasi-equilibrium with two reservoirs via ballistic transport: for example, particles injected from the source share the same distribution as the source.  Let $x_1$ denote the coordinate along which the potential difference is applied across the source and drain reservoirs. The generalized current injected from reservoir 1 flowing in the positive $x_1$ direction is given by $J_1 = g L^{-d} \sum (v_{g1}(\mathbf{k}))^a E^{b} f_{\pm}(E)$ in each valley of the dispersion.  Here $v_{g1}(\mathbf{k}) = (\hbar^{-1} \nabla_{\bf k}E ) \cdot \hat{x_1}$ is the group velocity projected along the $x_1$ coordinate, $a$ and $b$ may be fractions or integers, and $g$ combines degeneracies (e.g. valley, spin, polarization) and physical constants (e.g. electron charge, mass). The sum runs over all $\mathbf{k}$ states in the dispersion such that $k_1>0$. Choice of exponents $a$ and $b$ of 0 or 1 describe scalar particle densities or vector current densities. The subscript in $J_i$ denotes the reservoir from which the current in injected. The net current flowing from reservoir $1$ to reservoir $2$ along the positive $x_1$ direction is $J_{net}=J_1 +  J_2$ for scalar densities (e.g. particle density or energy density) and $J_{net}=J_1 -  J_2$ for vector current densities (e.g. particle current densities or energy current densities).  The parameters $\beta$ and $\mu$ in $f_{\pm}$ are dictated by the respective reservoirs, and the group velocity neglects Berry-phase contributions.

{\bf Main Result:} The generalized current $J_1$ can be recast as linear combination of sums of the type $I_{d,t}^{u,s} = \sum_{\Omega_{k_1}} k_1^u E^s f_{\pm}(E)$ that run over grid points in the $d-$dimensional hemisphere $\Omega_{k_1}$ for $k_1 \geq 0$. This converts to the integral

\begin{equation}
    I_{d,t}^{u,s} = \int_{k_1 = 0}^{\infty} \int_{k_2=-\infty}^{\infty} ... \int_{k_d=-\infty}^{\infty} \frac{\mathrm{d}k_1 \mathrm{d}k_2...\mathrm{d}k_d}{(\frac{2 \pi}{L})^d} k_1^u E^s f_{\pm}(E).
\end{equation}

Substituting $\alpha_i k_i\to k_i$ and splitting off $k_1$ using $k_0^2 = k_1^2 + \tilde{k}^2$ where $\tilde{k}^2 = k_2^2+...+k_d^2$, then passing into spherical coordinates $\mathrm{d}^{d-1}\tilde{k} = S_{d-2} \tilde{k}^{d-2} \mathrm{d}\tilde{k} $ where $S_{d-1} = 2 \pi^{d/2} / \Gamma(d/2)$ and $\Gamma(...)$ is the Gamma function, this becomes

\begin{equation}
    I_{d,t}^{u,s} = \frac{ S_{d-2} }{ \frac{(2 \pi)^d\alpha_1^{u}(\prod_{i=1}^d\alpha_i)}{L^d} }  \int_{0}^{\infty} \mathrm{d}k_1 k_1^u \int_{k_1}^{\infty} \mathrm{d}k_0 \hspace*{0.5mm}  k_0  (k_0^2 - k_1^2)^{\frac{d-3}{2}} E^s f_{\pm}(E),
    \label{general_integral_step2}
\end{equation}

which upon switching the order of integration evaluates to the exact closed form

\begin{equation}
    I_{d,t}^{u,s} = \left(\frac{L}{\lambda_{dB}}\right)^{d} \cdot \frac{1}{\beta^s \left(\frac{\lambda_{dB_1}}{2 \sqrt{\pi}}\right)^{u}} \cdot \frac{\Gamma(\frac{u+1}{2}) \Gamma(s+\frac{d+u}{t})}{t \sqrt{\pi} \Gamma(\frac{d+u}{2})} \cdot F^{\pm}_{s + \frac{d+u}{t} - 1}(\eta),
    \label{general_integral}
\end{equation}

where $\lambda_{dB}^d = \lambda_{dB_1} ... \lambda_{dB_d} = (4\pi)^{d/2}(\alpha_1 \alpha_2 ... \alpha_d) \beta^{d/t}$, and $\lambda_{dB_i} = \sqrt{4\pi} \alpha_i \beta^{1/t}$ is the generalized anisotropic thermal de-Broglie wavelength in the direction $i$ that characterizes the spatial spread of the wavepacket carrying the current.  For example, $\lambda_{dB_i} = h / \sqrt{2 \pi m_i k_b T}$ for parabolic ($t=2$) and $\lambda_{dB_i} =h v_i/\sqrt{\pi} k_b T$ for Dirac-like ($t=1$) dispersion.  $F^{\pm}_{j}(\eta) = \frac{1}{\Gamma(j+1)}  \int_{0}^{\infty}\,\mathrm{d}x \frac{x^{j}}{ \exp{[x - \eta]} \pm 1 }$ is the Fermi-Dirac or Bose-Einstein integral \cite{Fermi}. Though Equation \ref{general_integral_step2} is not defined for $d=1$, Equation \ref{general_integral} holds for all $d$. 

The generalized current in terms of Equation \ref{general_integral} therefore is $J_1 = g L^{-d} \sum (v_{g1}(\mathbf{k}))^a E^{b} f_{\pm}(E) = g/L^{d}(t\alpha_1^2/\hbar)^a I_{d,t}^{a, b+a - \frac{2a}{t}}$, which takes the compact form 

\begin{equation}
    J_{d,t}^{a,b} = g\cdot\frac{1}{\lambda_{dB}^{d} \beta^b} \cdot \left(\frac{\lambda_{dB_1}}{h \beta}\right)^{a} \cdot C_{d,t}^{a,b} \cdot F_{j}^{\pm}(\eta),
    \label{general_current}
\end{equation}

which is the main result of this work. $J_{d,t}^{a,b}$ is an explicit closed formula for $J_1$. Physically, this is the desired single expression for the density and current of particles, momentum, or heat, carried by both Fermions and Bosons flowing in the $x_1$ direction and injected from reservoir 1.  Here $a$ and $b$ are the exponents of the velocity and energy. $j=a+b+r-1$, $r=\frac{d-a}{t}$, and $C_{d,t}^{a,b} = \frac{ \Gamma(\frac{1+a}{2}) \Gamma(j+1 ) }{ (t\sqrt{\pi})^{1-a} \Gamma( \frac{a+d}{2} ) }$ are constants that depend in a simple and compact way on the dimension $d$, bandstructure type $t$, and type of current (e.g. particle, momentum, heat, etc) via the whole numbers $(a,b)$. The four numbers $(d, t, a, b)$ via Equation \ref{general_current} thus yield all currents. 

The interpretation of Equation 4 as a generalized current density becomes transparent by identifying it as a product of the following quantities: $g$, which represents physical constants and/or degeneracies, $1/(\lambda_{dB}^{d} \beta^b)$ which is dimensionally the (energy)$^b$/volume, $({\lambda_{dB_1}}/{h \beta})^{a}$ which is dimensionally the (velocity)$^a$, $C_{d,t}^{a,b}$ which is a dimensionless constant of order 1 for choices of $\{d,t,a,b\}$, and the dimensionless Fermi-Dirac or the Bose-Einstein integral $F_{j}^{\pm}(\eta)$.    Since this is a new general formulation for ballistic transport, we expect it to both unify previously known transport phenomena in new light, and also predict new phenomena. We highlight both aspects in the rest of this work. 

{\bf Low Temperature Asymptotics:} To highlight the utility of the unified formalism, we first explore the low temperature limits of generalized Fermion and Boson currents.  For example, the ballistic charge current $(a=1,b=0)$ in parabolic bands $(t=2)$ in $d-$dimensions for Fermions is obtained by choosing $F^{+}$ in Equation \ref{general_current} as $J_{d,2}^{1,0}= q(\frac{k_bT}{h})(\frac{2\pi m k_bT}{h^2})^{\frac{d-1}{2}}F_{\frac{d-1}{2}}^{+}(\frac{\mu}{k_bT})$, which in the limit of a highly degenerate Fermion distribution is $J_{d,2}^{1,0}(T \ll \mu / k_b )=\frac{q}{h^d}(2\pi m)^{\frac{d-1}{2}}\frac{\mu^{\frac{d+1}{2}}}{\Gamma(\frac{d+1}{2})}\left[1+\frac{\pi^2}{6}\frac{\Gamma(\frac{d+1}{2})}{\Gamma(\frac{d-1}{2})}\left(\frac{k_bT}{\mu}\right)^2\right]$, indicating a $T^2$ dependence.  While yielding the transport coefficients explicitly for different dimensions $d$, the above low temperature limit of ballistic charge current shows that this $T^2$ dependence is independent of the dimensions.  The independence of the $J \sim T^2$ dependence actually is seen to extend beyond the dimensionality, to other ballistic currents, which include heat or energy currents with a general $(a,b)$ and also to other dispersions (all $t$), because when expanded at low temperatures for fermions for $\mu >0$ up to $\mathcal{O}(T^2)$, Equation \ref{general_current} gives:
\begin{equation}
     J_{d,t}^{a,b}(T\ll\frac{\mu}{k_b}) = \frac{g (4\pi)^\frac{a-d}{2}\alpha_1^a C_{d,t}^{a,b}}{(\alpha_1 \alpha_2 ... \alpha_d)h^a}\frac{\mu^{j+1}}{\Gamma(j+1)}\left[1+\frac{\pi^2}{6}j\left(\frac{k_bT}{\mu}\right)^2\right]
\end{equation}

which guarantees the same temperature dependence for all dimensions $d$, as well as for all currents $(a,b)$ and types of bandstructures.

Unlike the `universal' $T^2$ dependence that results from the Sommerfeld expansion for all Fermion currents in the degenerate limit, that of Bosons depends on the dimensions, bandstructure, and the type of current.  Bose-Einstein statistics enforces $\eta \to 0$ as $T\to 0 $ for all dimensions \cite{cowan2019chemical}.  In the degenerate limit the generalized Bosonic current obtained from Equation \ref{general_current} is

\begin{equation}
    J_{d,t}^{a,b}(T\to0) = \frac{g (4\pi)^\frac{a-d}{2}\alpha_1^a C_{d,t}^{a,b}}{(\alpha_1 \alpha_2 ... \alpha_d)h^a} (k_bT)^{j+1}\zeta(j+1),\label{boson_lowtemp}
\end{equation}
where $\zeta(...)$ is the zeta-function.  As an example the thermal energy density $(a=0, b=1)$ stored in long-wavelength acoustic phonons with a linear dispersion ($t=1$) in a $d-$dimensional crystal is $J_{d,1}^{0,1}(T\to0) \propto T^{d+1}$, the specific case of $J_{d,t}^{a,b}(T\to0) \propto T^{a+b+\frac{d-a}{t}}$, which leads to a heat capacity $\sim T^d$.  We now remove the low temperature restriction to systematically illustrate with assorted examples the versatility of the new formulation in unifying the treatment of several disparate physical phenomena across dispersions and dimensions, and in predicting new phenomena. \\

{\bf I: Particle Densities ($a=0$, $b=0$):} From Equation \ref{general_current} the generalized particle density for various statistics, dispersions, and dimensions is obtained with $a=0,b=0$:

\begin{equation}
   n_{d,t} = 2 J_{d,t}^{0,0} = \frac{2g}{\lambda_{dB}^d} \cdot \frac{\Gamma(\frac{d}{t})}{t \Gamma(\frac{d}{2})} \cdot F_{\frac{d}{t}-1}^{\pm} (\eta).
   \label{general_density}
\end{equation}

The number density of photons of $g=2$ polarizations in thermal equilibrium with a radiation source at temperature $T$ is obtained using $F_{\frac{d}{t}-1}^{-} (0)$ in Equation \ref{general_density}. The chemical potential $\mu = 0$ for photons which are bosons whose particle number is not conserved in thermodynamic equilibrium with matter at temperature $T$.  In $d=3$ it is $2J_{3,1}^{0,0} = 16 \pi \zeta(3) ( \frac{ k_b T }{h c} )^3$ where $c$ is the speed of light, and in $d=2$ is $2J_{2,1}^{0,0}= 2 \pi^2 (\frac{ k_b T }{ h  c })^2$.  Because the photon has a positive branch dispersion, no energy gap, and Bose-Einstein statistics, no mass action law exists unlike for electrons and holes in semiconductors.

For a $t=2$ parabolic conduction band energy dispersion with $E = E_c + \sum_{i=1}^{d} (\alpha_{ci} k_i)^2$ with spin degeneracy $g_s=2$, valley degeneracy $g_c$ and the $+$ve sign for fermions, Equation \ref{general_density} gives the generalized volume density of electrons in $d$-dimensions $n_d= 2 J_{d,2}^{0,0} = N_c^{d} F^{+}_{(d/2)-1}[(\mu-E_c)/k_bT]$ where the band-edge density of states $N_c^d = 2g_c/\lambda_{dBc}^{d}$ is twice the inverse of the conduction band edge thermal de Broglie volume \cite{ashcroft1976solid, harrison1980solid}. The equivalent $d-$dimensional distribution for the valence band $E = E_v - \sum_{i=1}^{d} (\alpha_{vi} k_i)^2$ is $p_d = N_v^{d} F^{+}_{(d/2)-1}[(E_{v}-\mu)/k_bT]$.  For an energy gap $E_c - E_v = E_g$, the $d-$dimensional mass-action law governing equilibrium carrier statistics is $n_d p_d = n_{id}^2 $ which is obtained with $n_{id} \approx \sqrt{N_c^d N_v^d} \exp [-E_g/2 k_b T] $.  

For $t=1$ with a conical energy dispersion $E = \hbar v_F |{\bf k}|$, the fermion density per valley is $n_{d}(\mu) = 2 J_{d,1}^{0,0} = (4/\lambda_{dB}^d)(\Gamma(d)/\Gamma(d/2))F^{+}_{d-1}(\mu/k_bT)$.  If the Fermi level is at the Dirac point $\mu = 0$ for metallic carbon nanotubes ($d=1$), monolayer graphene ($d=2$), and HgCdTe ($d=3$), the intrinsic thermally generated electron density in each valley is $n_{di} = \frac{4}{(2\sqrt{\pi})^d} ( \frac{k_b T}{\hbar v_F})^d(\frac{\Gamma(d)}{\Gamma(d/2)}) \cdot F^{+}_{d-1}(0)$, varying with temperature as $n_i \sim T^d$ in $d-$dimensions.  This sets the lowest carrier density (and hence highest electrical resistivity) that may be reached in such materials at any temperature. For $E = \pm \hbar v_F |{\bf k}|$ where two cones touch, the sum of electron and hole densities is $n_{d}(+\mu)+n_{d}(-\mu)$, resulting in a corresponding mass-action law for Dirac dispersions.  The temperature dependence of the intrinsic electron/hole densities for conical bandstructure is therefore identical to the density of photons.\\

{\bf II: Energy Densities ($a=0$, $b=1$):}  The volume density of energy stored in a photon field in equilibrium with a radiation source of temperature $T$ is $2J_{d,1}^{0,1}$, which for $d=3$ is $\frac{4 \pi^5 (k_b T)^4}{(h c)^3}$, with corresponding results for other dimensions.  For long-wavelength acoustic phonons, the thermal energy stored in a solid is similarly obtained by choosing $g=1$ for each branch of sound velocity $v_s$ via $\alpha_i = \hbar v_s$, with $t=1$ and $a=0,b=1$. This gives the thermal energy density $2J_{3,1}^{0,1} = (\frac{4 \pi^5 k_b^4}{15 h^3 v_s^3}) T^4$, and a heat capacity per atomic density $n$ of $\frac{ C_v }{n} = 2 \partial J_{3,1}^{0,1} / \partial T = (\frac{16 \pi^5 k_b^4}{15 h^3 v_s^3}) T^3$, the $T\rightarrow 0$ limit of the Debye-$T^3$ law \cite{Debye1912, ashcroft1976solid}. 

\begin{figure*}[!htbp]
\includegraphics[scale=0.29]{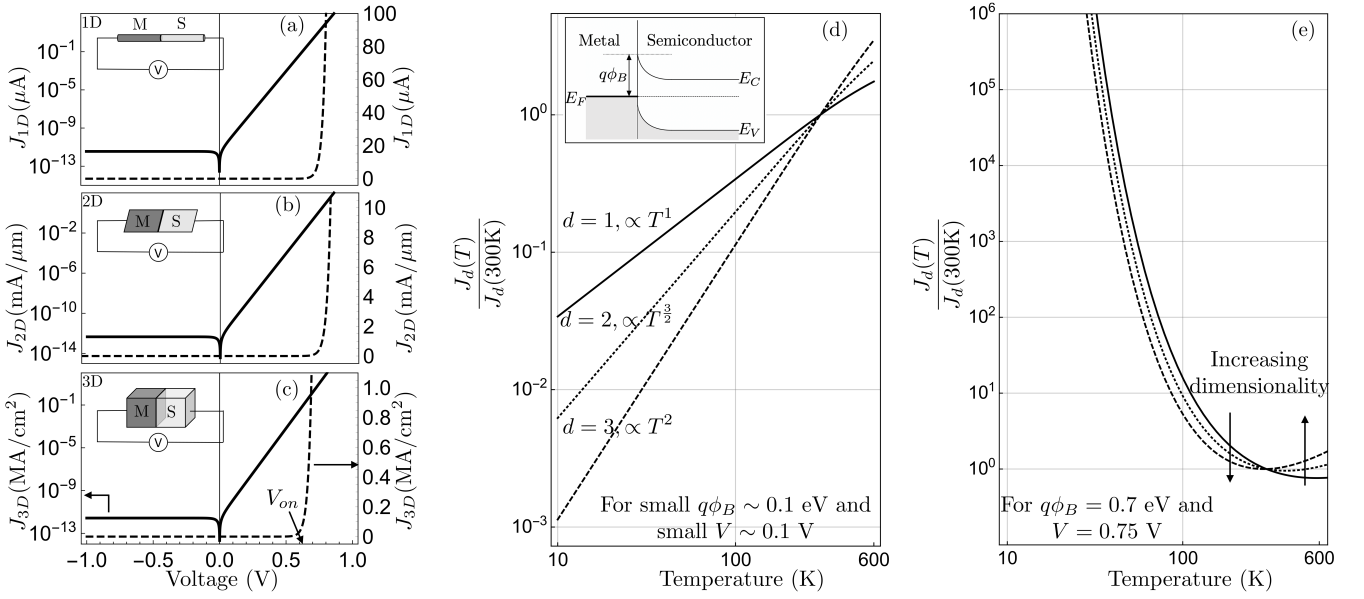}
\caption{\label{Schottky}(a), (b) and (c) are representative $300$ K $J - V$ characteristics of Metal (M) - Semiconductor (S) Schottky junctions in 1D, 2D and 3D respectively with $q\phi_B = 0.7$ eV where the semiconductor has parabolic dispersion ($t=2$). The solid curve is the logarithmic scale plot with axis on the left and the dashed curve is the linear scale plot with axis on the right. (d) ${J_d(T)}/{J_d(300 \text{ K})}$ vs. temperature for a small barrier ($q\phi_B\sim0.1$ eV) and small positive bias ($\sim0.1$  V) showing a $T^{\frac{d+1}{2}}$ dependence for $T<300 \text{ K}$. (e) Temperature dependence of ${J_d(T)}/{J_d(300 \text{ K})}$ converges for different dimensions for appreciable barrier heights and voltages.}
\end{figure*}
Because $J_{d,t}^{0,0}$ is the particle density and $J_{d,t}^{0,1}$ is the energy density, their ratio 

\begin{equation}
    u_{d,t} = \frac{J_{d,t}^{0,1}}{J_{d,t}^{0,0}} = \frac{d}{t} k_b T \frac{ F^{\pm}_{\frac{d}{t}}(\eta) }{ F^{\pm}_{\frac{d}{t}-1}(\eta) } \approx d\times \frac{k_b T}{t}
\end{equation}

is the generalized law of the equipartition of energy. For $-\eta>>1$, the Boltzmann approximation is valid for both Fermions and Bosons.  For particles in $d-$dimensions with mass and $t=2$, there is $k_b T/2$ energy per each dimension. For linear dispersion ($t=1$) on the other hand, there is $k_bT$ energy per each dimension as identified by Tolman \cite{tolman1979principles} in the relativistic limit and investigated further for other dispersions \cite{Turner1976, Buchdahl1984}. 

For degenerate fermions characterized by $\eta\gg+1$, the equilibrium average energy is $u_d \approx \mu d/(d+t)$ and the resulting electronic specific heat $c_v = \partial J_{d,t}^{0,1}/\partial T= \frac{g \pi^2 d}{3t^2}{\frac{k_bT}{\mu}} n_d k_b $ if $d/t\neq1$ and $c_v= \frac{g \pi^2 }{3}{\frac{k_bT}{\mu}} n_d k_b $ if $d/t=1$ \cite{tolman1979principles, landau2013course}.  For example, for electrons in metals with $d=3$, $g=2$, and $t=2$, $c_v = \frac{\pi^2}{2} \frac{k_b T}{\mu} n_{3d} k_b$, and for degenerately doped graphene with $d=2$, $g=4$ and $t=1$ is $c_v = \frac{8\pi^2}{3} \frac{k_b T}{\mu} n_{2d} k_b $. \\

{\bf III: Ballistic Charge Currents ($a=1$, $b=0$):} Suppose a solid with electronic bandstructure valleys of the types of Table \ref{tab:dispersions} is connected to two reservoirs held at the dimensionless potentials $\eta_1$ and $\eta_2$.  By setting $g=2q$ where $q$ is the electron charge of spin degeneracy $=2$, $a=1, b = 0$, while using $f^{+}$ for Fermions, Equation 4 yields the charge current density for each valley in quasi-equilibrium with the source reservoir:

\begin{equation}
    J_1 = J_{d,t}^{1,0} = \underbrace{\frac{2q^2}{h} \frac{\lambda_{dB_1}}{\lambda_{dB}^d} \frac{\Gamma(1+r)}{\Gamma(\frac{d+1}{2})} \frac{k_bT}{q} }_{J_0}F_{r}^{+}(\eta_1),
    \label{gen_ballistic_current}
\end{equation}
where $r=(d-1)/t$. The difference $J_{\text{net}} = J_1 - J_2 = J_0 [F^{+}_{r}(\eta_1) - F^{+}_{r}(\eta_2)]$ is the net macroscopic current, where the characteristic $J_0$ depends on $t,d$ and $\lambda_{dB}$, and is independent of the potential difference across the terminals.

The generalized form enables direct computation of ballistic currents in diodes and transistors of various dimensions and bandstructures. Applying Equation \ref{gen_ballistic_current} to a Schottky diode of electron barrier height $q\phi_b$ between a metal and a semiconductor with anisotropic bandstructure of dispersion type $t=2$ yields a generalized current density $\propto [F_{\frac{d-1}{t}}^{+}(\eta_1) - F_{\frac{d-1}{t}}^{+}(\eta_2)]$:

\begin{equation}
    J_{\text{schottky}} \approx \underbrace{ \frac{2q (2\pi m_e)^{\frac{d-1}{2}} k_b^{\frac{d+1}{2}}}{h^d} }_{A_{d,2}}  f_d  T^{\frac{d+1}{2}}  e^{-\frac{q\phi_b}{k_b T}}  (e^{\frac{qV}{k_b T}} - 1) \label{schottky}
\end{equation}

for $\beta(\eta_1 - \eta_2) = q V$ in the limit of $-\eta_1, -\eta_2 >>1$ as is typically the case in experiments. The case for $d=3$ was first derived by \citeauthor{bethe1942radiation} \cite{bethe1942radiation}; $A_{d,2}$ is the $d-$dimensional Richardson coefficient, and the dimensionless form factor  $f_d=\prod_{i\neq1} \sqrt{m_i / m_e}$ accounts for bandstructure anisotropy by excluding the mass component in the direction of transport. For 3D Silicon which has 6 valleys of the type $E = E_{c} + \frac{\hbar^2}{2}( \frac{k_x^2}{m_l} + \frac{k_y^2}{m_t} + \frac{k_z^2}{m_t} )$ along the 100 axis in $k$-space, the form factor for $d=3$ current along the 100 axis is $f_3 = \frac{2 m_t + 4\sqrt{m_l m_t}}{m_e}$ \cite{sze2006physics, crowell1965richardson} where the form factor is obtained from the 100 projections for each of the 6 valleys. The characteristic $J_0$ of Equation \ref{gen_ballistic_current} is the reverse saturation current density in $d$-dimensions for the diode relation given by Equation \ref{schottky}.  The formulation presented here therefore generalizes and extends the recent work of \citeauthor{Schottkyuniversal} \cite{Schottkyuniversal} which found that the lateral 2D Schottky reverse saturation current scales universally with temperature as $\ln(J_0/T^{3/2})\propto -1/T$. This result is extended to d-dimensional ballistic Schottky diodes using our formulation for $J_0$ in the generalized Richardson formula:
\begin{equation}
    \ln\left(\frac{J_{0}}{T^{\frac{d+1}{2}}}\right) = {\ln(A_{d,2}f)}-\frac{q\phi_b}{k_b T},
\end{equation}
yielding for $d=2$ the $\ln(J_0/T^{3/2})\propto -1/T$ dependence of 2D lateral Schottky heterojunctions.
For example, a lateral monolayer NbSe$_2$/WSe$_2$ junction forms a 2D-2D ballistic Schottky diode for which the current is $J_{\text{schottky}} \approx A_{2,2} \sqrt{\frac{m^{\star}}{m_e}} T^{\frac{3}{2}} e^{-\frac{q\phi_b}{k_b T}} \cdot (e^{\frac{qV}{k_b T}} - 1)$ for an isotropic 2D bandstructure. The ballistic current-voltage characteristics of Schottky diodes in $d$-dimensions calculated from the unified formula is shown in Fig.~\ref{Schottky} at $300$K. The formulation indicates the ranges of barrier heights and voltages in which the signature of the dimensionality should be imprinted in the variation of the ballistic current with temperature, and therefore experimentally measurable.

Equation \ref{gen_ballistic_current} also applies for ballistic electron transport in 2-terminal resistors, or 3-terminal field-effect transistors (FETs). For example, for a 2D electron gas channel with $d=2$ and bandstructure type $t=2$, the current per unit width per each valley is $J = \frac{2q^2}{h} \frac{1}{\lambda_{dB}} \frac{k_b T}{q} [ F^{+}_{1/2}(\eta_1) - F^{+}_{1/2}(\eta_2) ]$, in Natori's form \cite{Natori1994}. For bandstructure type $t=1$ and $d=2$ encountered in monolayer graphene or surface-bands of topological insulators, the current is  $J = \frac{2q^2}{h} \frac{2}{\sqrt{\pi}\lambda_{dB}}  \frac{k_b T}{q} [ F^{+}_{1}(\eta_1) - F^{+}_{1}(\eta_2) ]$. The 1D ballistic current per valley for $d=1$ is $J = \frac{2q^2}{h} \frac{k_b T}{q} \ln( \frac{ 1+ e^{\eta_1}}{1+ e^{\eta_2}} )$, which in the limit $\eta_1, \eta_2 >> +1$ typically encountered in experiments reduces to the Landauer limit \cite{Landauer_1989} given by $J = \frac{2q^2}{h} V$, indicating the conductance $J/V$ is quantized to $2q^2/h$ regardless of the type of bandstructure.  For ballistic currents for $t=2$, simultaneously fixing the total $d-$dimensional fermionic density $n_d = J_{d,t}^{0,0}(\eta_1) + J_{d,t}^{0,0}(\eta_2)$ (say via capacitive gate control) requires a self-consistent solution for $\eta_1$ and $\eta_2$ for charge and current, resulting in the saturation of the ballistic current beyond a certain voltage difference between the source and drain.  This is the hallmark of ballistic transistors that provide electronic gain for signal amplification, and switching for digital logic.\\


{\bf IV: Ballistic Heat Currents ($a=1$, $b=1$):} The heat current density is obtained directly from the entropy in the ballistic case using a Landauer approach (see for example \cite{Sivan1986}) or in the scattering-limited diffusive case using the  Boltzmann approach in the relaxation-time approximation (see for example \cite{smith1967electronic}). The ballistic heat current from an electrode is $Q = g L^{-d} \sum v_{g1}(k) (E-\mu) f_{\pm}(E)$, where $\mu$ is the chemical potential and $T$ the temperature of that electrode.  The generalized ballistic heat current density in quasi-equilibrium with the source reservoir is then obtained from Equation \ref{general_current} as $Q_1= J^{1,1}_{d,t}-\mu_1 J^{1,0}_{d,t}$:

\begin{equation}
    Q_1= \frac{gk_b^2}{h} \cdot \frac{\Gamma(r+1)}{\Gamma(\frac{d+1}{2})}\cdot\frac{\lambda_{dB_1}}{\lambda_{dB}^d}\cdot T^2[(1+r)F_{r+1}^{\pm}(\eta_1) - \eta_1 F_{r}^{\pm}(\eta_1)],
    \label{general_heat_current}
\end{equation}

and the net heat current density is $Q = Q_1 - Q_2$.

Since $\mu = 0$ for bosons whose particle number is not conserved, for $t=1$ and $v_i = c$ the net heat current with $f_{-}$ becomes

\begin{equation}
    Q_1-Q_2= \frac{g \pi^{\frac{d-1}{2}} k_b^{d+1}}{h^d c^{d-1}} \cdot \frac{\Gamma(d+1)}{\Gamma(\frac{d+1}{2})}\cdot F_{d}^{-}(0) \cdot [T_1^{d+1} - T_2^{d+1}],
    \label{boson_linear_heat_current}
\end{equation}

which is a generalized $d$-dimensional radiative cooling law. For a blackbody source at temperature $T_1=T$ radiating in $d=3$ dimensions and $g=2$ polarizations, Equation \ref{boson_linear_heat_current} yields $Q = (\frac{2 \pi^5 k_b^4}{15 c^2 h^3}) T^4$. This is the Stefan-Boltzmann radiation law \cite{Boltzmann1884, planck2013theory}, a spectral integral over the Planck blackbody radiation density in the photon field. The corresponding currents for blackbody radiators in $d=2$ is $J_{2,1}^{1,1} = (\frac{8 \zeta(3) k_b^3}{c h^2}) T^3$ and $d=1$ is $J_{1,1}^{1,1} = (\frac{\pi^2 k_b^2}{3 h}) T^2$.  The case of $d=1$ is special since it does not depend on the speed of light; indeed it is independent of the energy dispersion altogether because the velocity cancels the density of states.  Identical behavior exists for phonons and electrons, as discussed next.

For each branch of acoustic phonons, Equation \ref{boson_linear_heat_current} also gives the ballistic heat current between electrodes, with the speed of light replaced by the corresponding sound velocity.  When the drain electrode is at $T_2=0$ K, the $d=1$ heat current by an acoustic phonon branch of polarization $g=1$ is $J_{1,1}^{1,1} = (\pi^2 k_b^2 / 6 h) T^2$, identical to the photon current per polarization.  Though the ballistic phonon heat currents depend on temperature non-linearly, for $T_2 = T_0$ and a slightly hotter source at $T_1 = T_0 + \Delta T$, the heat current is 

\begin{equation}
    Q \approx \frac{g \pi^{\frac{d-1}{2}} k_b^{d+1}}{h^d v^{d-1}} \cdot \frac{\Gamma(d+2)}{\Gamma(\frac{d+1}{2})}\cdot F_{d}^{-}(0) \cdot T_0^d \cdot \Delta T,
\end{equation}

which is linear in temperature difference $Q = G \Delta T$. For $d=1$ the thermal conductance quantum $G_{0} =\pi^2 k_b^2T /( 3h )$ is obtained.  This was theoretically anticipated \cite{Thermalquantum1998PRL, angelescu1998heat} and subsequently experimentally observed \cite{Nature2000Thermalquantum}.


\begin{figure*}[!htbp]
\includegraphics[scale=0.4]{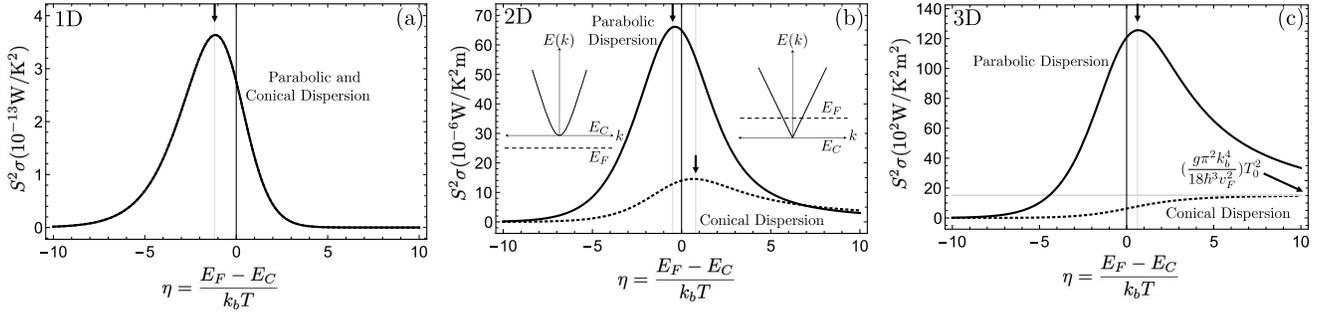}
\caption{\label{PF}Ballistic Power factor in different dimensions plotted as a function of the Fermi level location via the dimensionless parameter $\eta= (E_F-E_C)/k_bT$ at $300$ K. The dotted curve is the power factor for materials with conical dispersion $E=\hbar v_F|\mathbf{k}|$ with $v_F=10^6 \text{ m/s}$ and solid curve is the power factor for parabolic dispersion with $E=\hbar^2 k^2/2m$ with $m=0.2m_e$. The arrows on the top show the $\eta$ where the power factor shows maximum for both dispersions. Inset of Fig. 3(b) shows the conical and parabolic $E-k$ band-structures with the position of $E_F$ for maximum power factor.}
\end{figure*}
Because for electrons $\mu \neq 0$, Equation \ref{general_heat_current} gives a heat current dependent non-linearly on both $\mu$ and $T$ of the source and drain reservoirs. For small differences, for $t=2$ dispersion and $d=1$ to leading order in $\eta = \beta \mu \gg 1 $ it is 

\begin{equation}
    Q \approx \frac{g k_b^2 \pi^2 }{ 6 h }(T_1^2 - T_2^2) - \frac{g}{2h}(\mu_1^2 - \mu_2^2),
\end{equation}

which when linearized around a temperature $T$ and $\mu_1=\mu_2$ gives the same heat conductance quantum $\pi^2 k_b^2 T/(3 h)$ per spin channel as for photons and phonons.  In spite of the cancellation of the group velocity and the density of states in $d=1$, the heat conductance quantum due to electrons derives from its Fermionic statistics, yet is identical to the heat conductance quantum of phonons {\em and} photons that follow Bosonic statistics. This strange similarity was recognized in \cite{GreinerPRLcomment1998,RegoPRLreply1998,Greiner1997}, and Haldane's fractional exclusion statistics \cite{Haldane1991FS, Wu1995FS} was invoked to explain its possible origin \cite{rego1999fractional}.  The similarity of the 1d energy conductance quantum as a physical quantity independent of bosonic or fermionic statistics arising in the formulation here is traced to the following identities connecting the Fermi-Dirac and Bose-Einstein integrals:
\begin{eqnarray}
   F_1^{-}(0) = \frac{\pi^2}{6} =&& \lim_{\eta\to\infty} [ F_1^{+} (\eta)-\frac{(F_0^{+} (\eta))^2}{2F_{-1}^{+}(\eta)} ] ,\nonumber\\
   = && \lim_{\eta\to\infty} [ F_1^{+} (\eta)-\eta F_0^{+} (\eta) +\frac{\eta^2}{2}F_{-1}^{+}(\eta)].
\end{eqnarray}
Unlike photons and phonons though, the electron chemical potential difference also drives an energy current, which is captured well in the generalized linear transport coefficients.



{\bf V: Linear Response Coefficients:} Linearizing the above exact generalized formulations for ballistic transport for small differences in the reservoir chemical potentials $\mu_1 - \mu_2 = \Delta \mu$ and temperatures $\beta_1 - \beta_2 = \Delta \beta$ brings correlations between particle and energy currents into sharper focus. Instead of linearizing the distribution function (e.g. see \cite{lundstrom2012near} for ballistic and diffusive thermoelectric coefficients), here the unified generalized currents embodied by various choices of $(a,b)$ in Equation \ref{general_current} are expanded to linear order $J_{d,t}^{a,b} \approx g_{\mu}^{a,b} \Delta \mu + g_{\beta}^{a,b} \Delta \beta$ around the average chemical potential $\mu_0 = (\mu_1+\mu_2)/2$ and the average temperature $T_0$ given by $\beta_0 = 1/k_bT_0= (\beta_1+\beta_2)/2$. The linear coefficients are directly obtained as $g_{\mu}^{a,b} = (\partial J_{d,t}^{a,b}/\partial \mu)|_{\mu=\mu_0}$ and $g_{\beta}^{a,b} = (\partial J_{d,t}^{a,b}/\partial \beta)|_{\beta=\beta_0}$ and mapped to the traditional forms $J = L_{11} \Delta V + L_{12} \Delta T$ and $Q = L_{21} \Delta V + L_{22} \Delta T$, where $J=J_{d,t}^{1,0}$ is the charge current density and $Q=J^{1,1}_{d,t}-\mu_0 J^{1,0}_{d,t}$ is the heat current density in the linear response regime. Instead of the coefficients $L_{ij}$, the generalized linear coefficients obtained in experiments are the resistivity $\rho = \sigma^{-1} = L_{11}^{-1}$, the Seebeck coefficient $S = L_{12}/L_{11}$, the Peltier coefficient $\Pi = L_{21}/L_{11}$ and the electronic thermal conductivity $\kappa = L_{22}-L_{12}L_{21}/L_{11}$. The ballistic linear response coefficients obtained from Equation \ref{general_current} are

\begin{eqnarray}
    \rho = && \sigma^{-1} = (\frac{g_0 q^2}{h} \cdot \frac{\lambda_{dB_1}}{\lambda_{dB}^d} \cdot \frac{\Gamma(r+1)}{\Gamma(\frac{d+1}{2})} \cdot F_{r-1}^{+}(\eta) )^{-1}, \nonumber \\
    S = && -\frac{k_b}{q} [\eta - (r+1)\frac{F_{r}^{+}(\eta)}{F_{r-1}^{+}(\eta)}], \nonumber \\
    \Pi = && S \cdot T_0, \text{ and }  \nonumber \\
    \kappa =&& \frac{g_0 k_b^2 T_0}{h} \cdot \frac{\lambda_{dB_1}}{\lambda_{dB}^d} \cdot \frac{\Gamma(r+1)}{\Gamma(\frac{d+1}{2})} \cdot \nonumber
    \\&&\qquad [(r+1)(r+2)F_{r+1}^{+}(\eta) - (r+1)^2 \frac{(F_r^+(\eta))^2}{F_{r-1}^{+}(\eta)}].
    \label{ballistic_linear_response_coefficients}
\end{eqnarray}

where $g_0$ is the product of spin and valley degeneracies, $\eta = \mu_0 \beta_0$, and $r=(d-1)/t$ generalizes the expressions for the several bandstructure types and dimensions.  A conceptual difference of the ballistic coefficients is that the diffusive coefficients represent local properties, whereas the ballistic ones represent terminal (or system) properties as discussed lucidly for $d=1$ by Butcher in \cite{butcher1990thermal}.  The quantization of both $\sigma$ and $\kappa$ in $d=1$ for $\eta\gg+1$ is explicit for all $t$ in Equation \ref{ballistic_linear_response_coefficients}.  The Onsager symmetry relation $\Pi = S T_0$ is seen to remain valid for the ballistic situation for all $d,t$.  The generalized Lorenz number $\mathcal{L}_{d,t} = \kappa/(\sigma T_0)$ obtained from Equations \ref{ballistic_linear_response_coefficients} goes to $\mathcal{L}_{d,t} \rightarrow \frac{\pi^2}{3} (\frac{k_b}{q})^2$ in the degenerate fermion limit of $\beta \mu \gg 1$ for all $d$ and $t$, highlighting the robustness of the Wiedemann-Franz law in the ballistic limit \cite{ziman1972principles, butcher1990thermal}.  In the non-degenerate limit of $-\beta \mu >>1$ relevant for semiconductors, $\mathcal{L}_{d,t} \rightarrow (\frac{d-1}{t} + 1)(\frac{k_b}{q})^2$.  
\begin{table*}
\caption{\label{tab:all_cases_fermions_bosons}Generalized ballistic currents in $d-$dimensions for Fermions ($+$) and Bosons ($-$)}
\begin{ruledtabular}
\begin{tabular}{ccccc}
\multicolumn{5}{c}{$J_{d,t}^{a,b}$, with $\eta=\frac{\mu}{k_b T}$. \hspace{3mm}  [$t=1$: $E=\hbar v_F | \mathbf{k} | $] \& [$t=2$: $E = \frac{ \hbar^2 \lvert\mathbf{k}\rvert^2 }{ 2m }$]. \hspace{3mm} $F^{\pm}_{j}(\eta) = \frac{1}{\Gamma(j+1)}  \int_{0}^{\infty}\,\mathrm{d}x \frac{x^{j}}{ \exp{[x - \eta]} \pm 1 }$,  and  $F_{0}^{\pm}(\eta) = \pm \ln{ [ 1 \pm e^{\eta}] }$. }\\ \hline 
      &  Particle Density & Energy Density & Particle Current & Heat Current  \\ 
    $(d,t)\downarrow$  \text{     } $(a,b)\rightarrow$  &  $(2)J_{d,t}^{0,0}$ & $(2)J_{d,t}^{0,1}$ & $J_{d,t}^{1,0}$ & $J_{d,t}^{1,1}-\mu J_{d,t}^{1,0}$  \\ \hline
                                    &   & & &   \\
    $(1,1)$   & $  2g (\frac{ k_b T}{h v_F} ) F_{0}^{\pm}(\eta)$  & $2g\frac{(k_b T)^2}{h v_F} F_{1}^{\pm}(\eta)$ & $g \frac{k_b T}{h} F_{0}^{\pm}(\eta)$ & $g\frac{(k_b T)^2}{h} \left(F_{1}^{\pm}(\eta)- \eta F_{0}^{\pm}(\eta)\right)$ \\
                            &   & & &   \\
    $(2,1)$   & $ 2 \pi g (\frac{  k_b T }{ hv_F })^2 F_{1}^{\pm}(\eta)$  & $4\pi g \frac{(k_b T)^3}{(h v_F)^2}F_{2}^{\pm}(\eta)$ & $2g v_F (\frac{ k_b T }{ h v_F })^2 F_{1}^{\pm}(\eta)$ & $2g\frac{(k_b T)^3}{h^2 v_F} \left(2F_{2}^{\pm}(\eta)-\eta F_{1}^{\pm}(\eta)\right)$ \\
                            &   & & &   \\
    $(3,1)$   & $8 \pi g  (\frac{  k_b T }{ hv_F })^3 F_{2}^{\pm}(\eta)$  & $24 \pi g \frac{(k_b T)^4}{(h v_F)^3}F_{3}^{\pm}(\eta)$ & $2 \pi g v_F (\frac{k_b T}{h v_F})^3 F_{2}^{\pm}(\eta)$ & $2 \pi g\frac{(k_b T)^4}{h^3 v_F^2}\left( 3F_{3}^{\pm}(\eta)-\eta F_{2}^{\pm}(\eta)\right)$ \\ 
                            &   & & &   \\\hline
                            &   & & &   \\
    $(1,2)$   & $ g( \frac{2 \pi m k_b T }{h^2} )^{\frac{1}{2}} F_{-\frac{1}{2}}^{\pm}(\eta)$  & $ g \frac{k_b T}{2} ( \frac{2 \pi m k_b T }{h^2} )^{\frac{1}{2}} F_{\frac{1}{2}}^{\pm}(\eta)$ & $ g \frac{k_b T}{h} F_{0}^{\pm}(\eta)$ & $g \frac{(k_bT)^2}{h} \left(F_{1}^{\pm}(\eta)-\eta F_{0}^{\pm}(\eta)\right)$ \\
                            &   & & &   \\
    $(2,2)$   & $ g( \frac{2 \pi m k_b T }{h^2} ) F_{0}^{\pm}(\eta)$  & $ g \frac{2k_b T}{2} (\frac{2 \pi m k_b T }{h^2} ) F_{1}^{\pm}(\eta)$ & $g \frac{k_b T}{h} (\frac{2 \pi m k_B T}{h^2})^{\frac{1}{2}} F_{\frac{1}{2}}^{\pm}(\eta)$ & $ g\frac{(k_bT)^2}{h} (\frac{2 \pi m k_b T}{h^2})^{\frac{1}{2}} \left(\frac{3}{2}F_{\frac{3}{2}}^{\pm}(\eta)-\eta F_{\frac{1}{2}}^{\pm}(\eta)\right)$ \\
                            &   & & &   \\
    $(3,2)$   & $ g( \frac{2 \pi m k_b T }{h^2} )^{\frac{3}{2}} F_{\frac{1}{2}}^{\pm}(\eta)$  & $ g \frac{3k_b T}{2} ( \frac{2 \pi m k_b T }{h^2} )^{\frac{3}{2}} F_{\frac{3}{2}}^{\pm}(\eta)$ & $g \frac{k_b T}{h} (\frac{2 \pi m k_B T}{h^2}) F_{1}^{\pm}(\eta)$ & $g \frac{(k_bT)^2}{h} (\frac{2 \pi m k_b T}{h^2})\left( 2 F_{2}^{\pm}(\eta)-\eta F_{1}^{\pm}(\eta)\right)$ \\
                            &   & & &   \\
\end{tabular}
\end{ruledtabular}
\end{table*}

The generalized formulation of Equation \ref{ballistic_linear_response_coefficients} brings a novel feature of the dependence of the ballistic power factor ($S^2\sigma$) on dimensions and band-structures into sharp focus, as highlighted in Figure \ref{PF}. Because the Seebeck coefficient $S \sim - \frac{k_b}{q} \cdot \frac{E_F - E_c}{k_b T}$ decreases with increasing $E_F$, whereas $\sigma$ increases with increasing $E_F$, conventional wisdom states that the power factor product $S^2 \sigma$ should exhibit a maximum somewhere near $E_F = E_c$. As Figure \ref{PF} shows, for all $d,t$ the ballistic themoelectric power factor $S^2 \sigma$ indeed shows a maximum near $\mu=0$, {\em except} for the $d=3, t=1$ conical electron energy dispersion. For this case, it increases monotonically with $\mu$ and saturates to $S^2 \sigma \rightarrow (\frac{g \pi^2 k_b^4}{18 \hbar^3 v_F^2})T_0^2$.  This behavior has neither been identified theoretically, nor observed experimentally in the past. This dependence of the power factor on the dimensionality warrants an experimental search for the monotonic increase with the Fermi level. Such behavior could potentially be observed in the bulk states of 3D topological Dirac semimetals such as $\textrm{Na}_3\textrm{Bi}$ \cite{Na3Bi} and $\textrm{Cd}_3\textrm{As}_2$ \cite{Cd3As2}. This prediction emerged from the ballistic transport study, and highlights an example of the value of the generalized $d-$dimensional formulation for various bandstructures that is achieved in this work. \\

{\bf Conclusions and Future Directions:} The generalized ballistic current expression obtained in Equation \ref{general_current} is found to be a versatile tool to compute and compare in a unified manner the particle and energy densities, charge and energy currents, thermoelectric coefficients and more for fermions and bosons of various energy dispersions. Such a compact formulation is well suited for optimization problems, in which the extrema of one or more densities, currents, transport coefficents, or their combinations need to be determined as a function of the dimensionality, type of dispersion, effective masses, wave velocities etc.  To facilitate such studies, the generalized ballistic currents $J_{d,t}^{a,b}$ for various $a,b$ are summarized in Table \ref{tab:all_cases_fermions_bosons}, and Table \ref{tab:all_cases_linear_coeff} shows the linear response coefficients.   

The energy dispersion types are not restricted to the specific cases of $t=1,2$ discussed, or to integers. The ballistic current expression may be extended for mixed dispersions of the tight-binding type $E = E_0 + 2t \cos{ka} \approx E_0 + 2t [ 1-(ka)^2/2 +(ka)^4/24 ... ]$ near band edges, and to those that involve $k_i k_j$ and $k_i^a + k_j^b$, as present in some realistic systems, and topologically non-trivial terms may be introduced. Extending the formulation to multi-terminal cases in the spirit of the Landauer–B{\"u}ttiker formalism \cite{buttiker1986PRL, datta1997electronic}, and especially for generalized nonlinear response in a magnetic field for various dimensions and dispersions is of high interest. So is exploring the various non-linear response predictions for ballistic electronic and thermoelectric transport phenomena.  Extension of this approach to ballistic particle and energy transport in hetero-dimensional situations (mixed $d$), and for mixed dispersions and statistics (e.g. plasmons or phonon-polaritons) is also suggested as future work.  The formulation is not limited to electrons, photons and phonons as discussed here, and is applicable to molecular systems that undergo ballistic motion.  Ballistic electron transport in condensed matter systems is seen primarily in nanoscale structures, which also have small numbers of particles, sometimes on the verge of failing the large number requirements on which traditional thermodynamic relations rest.  The implications of recently revealed non-equilibrium thermodynamics equalities in nanoscale systems and on fluctuations of the densities, energies, and currents discussed here are therefore of significant theoretical and practical interest \cite{Jarzynski1997PRL, Crooks1999PRE}.

\begin{acknowledgements}
This work was supported in part by the National Science Foundation under the NewLAW EFRI (1741694) and the E2CDA (ECCS 1740286) programs. The authors sincerely thank Drs.\ Farhan Rana, Francesco Monticone, Jacob Khurgin, Huili (Grace) Xing and Menyoung Lee for reading and commenting on the contents of this manuscript.  They are grateful to an anonymous reviewer for detailed comments that resulted in several improvements in the notation and clarity of this work.
\end{acknowledgements}


\begin{table*}
\caption{\label{tab:all_cases_linear_coeff}Generalized ballistic linear response coefficients in $d-$dimensions}
\begin{ruledtabular}
\begin{tabular}{cccc}
\multicolumn{4}{c}{$\eta=\frac{\mu_0}{k_b T_0}$ and $g_0=$ (spin degeneracy)$\times$(valley degeneracy).  \hspace{3mm} Note: The Peltier Coefficient $\Pi=S.T_0$ by the Onsager relation.}\\ \hline 
      &  Resistivity & Seebeck Coefficient & Thermal Conductivity  \\ 
    $(d,t)\downarrow$   &  $\rho=\sigma^{-1}$ & $S$& $\kappa$  \\ \hline
                                    &   & &   \\
    $(1,1)$   & $ \left(\frac{g_0q^2}{h}\frac{1}{1+e^{-\eta}}\right)^{-1}$  & $-\frac{k_b}{q}\left(\eta-(1+e^{-\eta})\ln[1+e^{\eta}]\right)$ & $\frac{g_0k_b^2 T_0}{h}\left( 2F_{1}^{+}(\eta)-(1+e^{-\eta})\ln^2[1+e^{\eta}]\right)$ \\
                            &   &  &   \\
    $(2,1)$   & $ \left(\frac{2g_0q^2}{h}(\frac{k_bT_0}{hv_F})\ln [1+e^{\eta}]\right)^{-1}$  & $-\frac{k_b}{q}\left(\eta-\frac{2F_{1}^{+}(\eta)}{\ln[1+e^{\eta}]}\right)$  & $4g_0k_b\frac{(k_b T_0)^2}{h^2 v_F}\left(3F_{2}^{+}(\eta)-\frac{2(F_{1}^{+}(\eta))^2}{\ln[1+e^{\eta}]}\right) $ \\
                            &   &  &   \\
    $(3,1)$   & $\left(\frac{2\pi g_0q^2}{h}(\frac{k_bT_0}{hv_F})^2 F_{1}^{+}(\eta)\right)^{-1}$  & $-\frac{k_b}{q}\left(\eta-\frac{3F_{2}^{+}(\eta)}{F_{1}^{+}(\eta)}\right)$  & $6 \pi g_0k_b\frac{(k_b T_0)^3}{h^3 v_F^2} \left(4F_{3}^{+}(\eta)-3\frac{(F_{2}^{+}(\eta))^2}{F_{1}^{+}(\eta)}\right)$ \\ 
                            &   &  &   \\\hline
                            &   &  &   \\
    $(1,2)$   & $\left(\frac{g_0q^2}{h}\frac{1}{1+e^{-\eta}}\right)^{-1}$  & $-\frac{k_b}{q}\left(\eta-(1+e^{-\eta})\ln[1+e^{\eta}]\right)$  & $\frac{g_0k_b^2 T_0}{h}\left( 2F_{1}^{+}(\eta)-(1+e^{-\eta})\ln^2[1+e^{\eta}]\right)$  \\
                            &   &  &   \\
    $(2,2)$   & $\left(\frac{g_0q^2}{h}(\frac{2\pi m k_bT_0}{h^2})^{\frac{1}{2}}F_{-\frac{1}{2}}^{+}(\eta)\right)^{-1}$  & $-\frac{k_b}{q}\left(\eta-\frac{3}{2}\frac{F_{\frac{1}{2}}^{+}(\eta)}{F_{-\frac{1}{2}}^{+}(\eta)}\right)$ & $\frac{3g_0}{4} \frac{k_b^2T_0}{h} (\frac{2 \pi m k_b T_0}{h^2})^{\frac{1}{2}} \left(5F_{\frac{3}{2}}^{+}(\eta)-3\frac{(F_{\frac{1}{2}}^{+}(\eta))^2}{F_{-\frac{1}{2}}^{+}(\eta)}\right)$ \\
                            &   &  &   \\
    $(3,2)$   & $\left(\frac{g_0q^2}{h}(\frac{2\pi m k_b T_0}{h^2})\ln[1+e^{\eta}]\right)^{-1}$  & $-\frac{k_b}{q}\left(\eta-\frac{2F_{1}^{+}(\eta)}{\ln[1+e^{\eta}]}\right)$ & $2g_0 \frac{k_b^2T_0}{h} (\frac{2 \pi m k_b T_0}{h^2}) \left(3F_{2}^{+}(\eta)-2\frac{(F_{1}^{+}(\eta))^2}{\ln[1+e^{\eta}]}\right)$ \\
                            &   &  &   \\
\end{tabular}
\end{ruledtabular}
\end{table*}

\bibliography{general_ballistic}

\begin{thebibliography}{39}
\bibitem [{Fer()}]{Fermi}%
  \BibitemOpen
  \bibinfo {title} {The asymptotic approximations of the {F}ermi-{D}irac
  {I}ntegral are: for {$\eta\ll-1$, ${F}_j^{+}(\eta) \approx e^{\eta}$ and for
  $\eta\gg1$, ${F}_j^{+}(\eta) \approx \eta^{j+1}/\Gamma(j+2)$. The $\eta=0$
  value of the Bose-Einstein Integral is ${F}_j^{-}(0) = \zeta(j+1)$, where
  $\zeta ( ... )$ is the Zeta-function.}}\BibitemShut {Stop}%
\end{thebibliography}

\end{document}